# A population of red candidate massive galaxies ~600 Myr after the Big Bang

Ivo Labbé[1], Pieter van Dokkum[2], Erica Nelson[3], Rachel Bezanson[4], Katherine A. Suess[5,6], Joel Leja[7,8,9], Gabriel Brammer[10], Katherine Whitaker[10,11], Elijah Mathews[7,8,9], Mauro Stefanon[12,13], Bingjie Wang[7,8,9]

[1]Centre for Astrophysics and Supercomputing, Swinburne University of Technology, Melbourne, VIC 3122, Australia
[2]Department of Astronomy, Yale University, New Haven, CT 06511, USA
[3]Department for Astrophysical and Planetary Science, University of Colorado, Boulder, CO 80309, USA
[4]Department of Physics and Astronomy and PITT PACC, University of Pittsburgh, Pittsburgh, PA 15260, USA
[5]Department of Astronomy and Astrophysics, University of California, Santa Cruz, 1156 High Street, Santa Cruz, CA 95064 USA
[6]Kavli Institute for Particle Astrophysics and Cosmology and Department of Physics, Stanford University, Stanford, CA 94305, USA
[7]Department of Astronomy & Astrophysics, The Pennsylvania State University, University Park, PA 16802, USA
[8]Institute for Computational & Data Sciences, The Pennsylvania State University, University Park, PA, USA
[9]Institute for Gravitation and the Cosmos, The Pennsylvania State University, University Park, PA 16802, USA
[10]Cosmic Dawn Center (DAWN), Niels Bohr Institute, University of Copenhagen, Jagtvej 128, København N, DK-2200, Denmark
[11] Department of Astronomy, University of Massachusetts, Amherst, MA 01003, USA
[12] Departament d'Astronomia i Astrofisica, Universitat de Valencia, C. Dr. Moliner 50, E-46100 Burjassot, Valencia, Spain
[13]Unidad Asociada CSIC "Grupo de Astrofisica Extragalactica y Cosmologi" (Instituto de Fisica de Cantabria - Universitat de Valencia)

**Galaxies with stellar masses as high as ~ $10^{11}$ solar masses have been identified[1–3] out to redshifts z ~ 6, approximately one billion years after the Big Bang. It has been difficult to find massive galaxies at even earlier times, as the Balmer break region, which is needed for accurate mass estimates, is redshifted to wavelengths beyond 2.5 μm. Here we make use of the 1-5 μm coverage of the *JWST* early release observations to search for intrinsically red galaxies in the first ≈ 750 million years of cosmic history. In the survey area, we find six candidate massive galaxies (stellar mass > $10^{10}$ solar masses) at 7.4 ≤ z ≤ 9.1, 500–700 Myr after the Big Bang, including one galaxy with a possible stellar mass of ~$10^{11}$ solar masses. If verified with spectroscopy, the stellar mass density in massive galaxies would be much higher than anticipated from previous studies based on rest-frame ultraviolet-selected samples.**

The galaxies were identified in the first observations of the *JWST* Cosmic Evolution Early Release Science (CEERS) program. This program obtained multi-band images at 1−5 μm with the Near Infrared Camera (NIRCam) in a "blank" field, chosen to overlap with existing *Hubble Space Telescope (HST)* imaging. The total area covered by these initial data is ≈ 40 arcmin[2]. The data were obtained from the MAST archive and reduced using the Grizli pipeline.[4] A catalog of sources was created, starting with detection in a deep combined

F277W+F356W+F444W image (see Methods for details). A total of 42,729 objects are in this parent catalog.

We selected candidate massive galaxies at high redshifts by identifying objects that have two redshifted breaks in their spectral energy distributions (SEDs), the $\lambda_{rest}$ = 1216 Å Lyman break and the $\lambda_{rest} \sim$ 3600 Å Balmer break. This selection ensures that the redshift probability distributions are well-constrained, have no secondary solutions at lower redshifts, and that we include galaxies that have potentially high mass-to-light ratios. Specifically, we require that: objects are not detected at optical wavelengths; blue in the near-infrared with F150W – F277W< 0.7; red at longer wavelengths with F277W – F444W > 1.0; brighter than F444W < 27 AB magnitude. After visual inspection to remove obvious artefacts (such as diffraction spikes), this selection produced 13 galaxies with the sought-for "double-break" spectral energy distributions. Next, redshifts and stellar masses were determined with three widely-used techniques, taking the contribution of strong emission lines to the rest-frame optical photometry explicitly into account.[5–15] We use the EAZY code[16] (with additional strong emission line templates), the Prospector-α framework[17], and five configurations of the Bagpipes SED-fitting code to explore systematics due to modeling assumptions. The seven individual mass and redshift measurements of the 13 galaxies are listed in the Methods section. We adopt fiducial masses and redshifts by taking the median value for each galaxy. We note that these masses and redshifts are not definitive and that all galaxies should be considered candidates.

As shown in Fig. 1 all 13 objects have photometric redshifts 6.5 < z < 9.1. Six of the 13 have fiducial masses > $10^{10}$ M$_\odot$ (Salpeter IMF) and multi-band images and spectral energy distributions of these galaxies are shown in Figs. 2 and 3. Their photometric redshifts range from z=7.4 to z=9.1. The model fits are generally excellent, and in several cases clearly demonstrate that rest-frame optical emission lines contribute to the continuum emission. These lines can be so strong in young galaxies that they can dominate the broad band fluxes redward of the location of the Balmer break,[6–8,14,18] and *Spitzer*/IRAC detections of optical continuum breaks in galaxies at z ≳ 5 have been challenging to interpret.[3, 5, 19–24] With *JWST*, this ambiguity is largely resolved owing to the dense wavelength coverage of the NIRCam filters and the inclusion of relatively narrow emission line-sensitive filter F410M,[25] which falls within the F444W band, although the uncertainties are such that alternative solutions with lower masses may exist[14]. The brightest galaxy in the sample, 38094, is at z = 7.5 and may have a mass that is as high as M$_* \approx$ 1×$10^{11}$ M$_\odot$, more massive than the present-day Milky Way. It has two nearby companions with a similar break in their optical to near-IR SEDs, suggesting that the galaxy may be in a group.

We place these results in context by comparing them to previous studies of the evolution of the galaxy mass function to z ~ 9. These studies are based on samples that were selected in the rest-frame UV using ultra-deep HST images, with *Spitzer*/IRAC photometry typically acting as a constraint on the rest-frame optical SED.[3, 15, 26–28] The bottom panel of Fig. 3 compares the average SED of the six candidate massive galaxies to the SEDs of *HST*-selected galaxies at similar redshifts. The galaxies we report here are much redder and the differences are not limited to one or two photometric bands: the entire SED is qualitatively different. This is the key result of our study: we show that galaxies can be robustly identified at z>7 with *JWST* that are intrinsically-redder than previous *HST*-selected samples at the same redshifts. It is likely that these galaxies also have much higher M/L ratios, but this needs to be confirmed with spectroscopy. We note that the new galaxies are very faint in the rest-frame

UV (median F150W~28 AB), and previous wide-field studies with HST and Spitzer[29] of individual galaxies did not reach the required depths to find this population.

The masses that we derive are intriguing in the context of previous studies. No candidate galaxies with $\log(M_*/M_\odot) > 10.5$ had been found before beyond $z \sim 7$, and no candidates with $\log(M_*/M_\odot) > 10$ had been found beyond $z \sim 8$. Furthermore, Schechter fits to the previous candidates predicted extremely low number densities of such galaxies at the highest redshifts.[3] This is shown by the lines in Fig. 4: the expected mass density in galaxies with $\log(M_*/M_\odot) > 10$ at $z \sim 9$ was $\sim 10^2 \, M_\odot \, \text{Mpc}^{-3}$, and the *total* previously derived stellar mass density, integrated over the range $8 < \log(M_*/M_\odot) < 12$, is less than $10^5 \, M_\odot \, \text{Mpc}^{-3}$. If confirmed, the *JWST*-selected objects would fall in a different region of Fig. 4, in the top right, as the *JWST*-derived fiducial mass densities are far higher than the expected values based on the UV-selected samples. The mass in galaxies with $\log(M_*/M_\odot) > 10$ would be a factor of $\sim 20$ higher at $z \sim 8$ and a factor of $\sim 1000$ higher at $z \sim 9$. The differences are even greater for $\log(M_*/M_\odot) > 10.5$.

We infer that the possible interpretation of these *JWST*-identified "optical break galaxies" falls between two extremes. If the redshifts and fiducial masses are correct, then the mass density in the most massive galaxies would exceed the *total* previously estimated mass density (integrated down to $M_* = 10^8 M_\odot$) by a factor of $\sim 2$ at $z \sim 8$ and by a factor of $\sim 5$ at $z \sim 9$. Unless the low mass samples are highly incomplete, the implication would be that most of the total stellar mass at $z = 8 - 9$ resides in the most massive galaxies. Although extreme, this is qualitatively consistent with the notion that the central regions of present-day massive elliptical galaxies host the oldest stars in the universe (together with globular clusters), and with the finding that by $z \sim 2$ the stars in the central regions of massive galaxies already make up $10\% - 20\%$ of the total stellar mass density at that redshift.[30] A more fundamental issue is that these stellar mass densities are difficult to realize in a standard LCDM cosmology, as pointed out by several recent studies.[31,32] Our fiducial mass densities push against the limit set by the number of available baryons in the most massive dark matter halos.

The other extreme interpretation is that all the fiducial masses are larger than the true masses by factors of >10-100. We use standard techniques and multiple methods to estimate the masses. Under certain assumptions for the dust attenuation law and stellar population age sampling (favoring young ages with strong emission lines), low masses can be produced (see Methods). This only occurs at specific redshifts (z=5.6, 6.9, 7.7, or about ~10% of the redshift range of the sample) where line-dominated and continuum-dominated models produce similar F410M-F444W colors. In addition, it is possible that techniques that have been calibrated with lower redshift objects[17] are not applicable. As an example, we do not include effects of exotic emission lines or bright active galactic nuclei (AGN)[14]. Part the sample is reported to be resolved in F200W[33] making significant contribution from AGN less likely, but faint, red AGN are possible and would be highly interesting in their own right, even if they could lead to changes in the masses.

It is perhaps most likely that the situation is in between these extremes, with some of the red colours reflecting exotic effects or AGN and others reflecting high M/L ratios. Future *JWST* NIRSpec spectroscopy can be used to measure accurate redshifts as well as the precise contributions of emission lines and to the observed photometry. With deeper data the stellar continuum emission can be detected directly for the brightest galaxies. Finally, dynamical masses are needed to test the hypothesis that our description of massive halo assembly in

LCDM is incomplete. It may be possible to measure the required kinematics with ALMA or from rotation curves with the NIRSpec IFU if the ionized gas is spatially extended.[30,31]

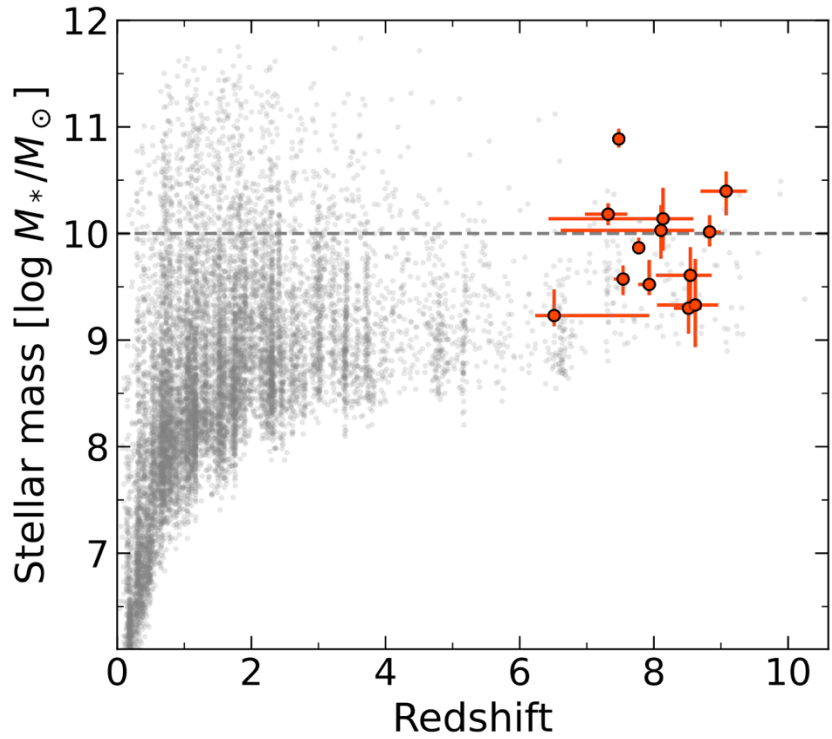

Figure 1: **Redshifts and tentative stellar masses of double-break selected galaxies**. Shown in gray circles are EAZY-determined redshifts and stellar masses using emission-line enhanced templates (Salpeter IMF) for objects with S/N> 8 in the F444W band. Fiducial redshifts and masses of the bright galaxies (F444W < 27 AB) that satisfy our double-break selection are shown by the large red symbols. Uncertainties are the 16$^{th}$-84$^{th}$ percentile of the posterior probability distribution. All galaxies have photometric redshifts $6.5 < z < 9.1$. Six galaxies are candidate massive galaxies with fiducial $M_* > 10^{10}$ $M_\odot$.

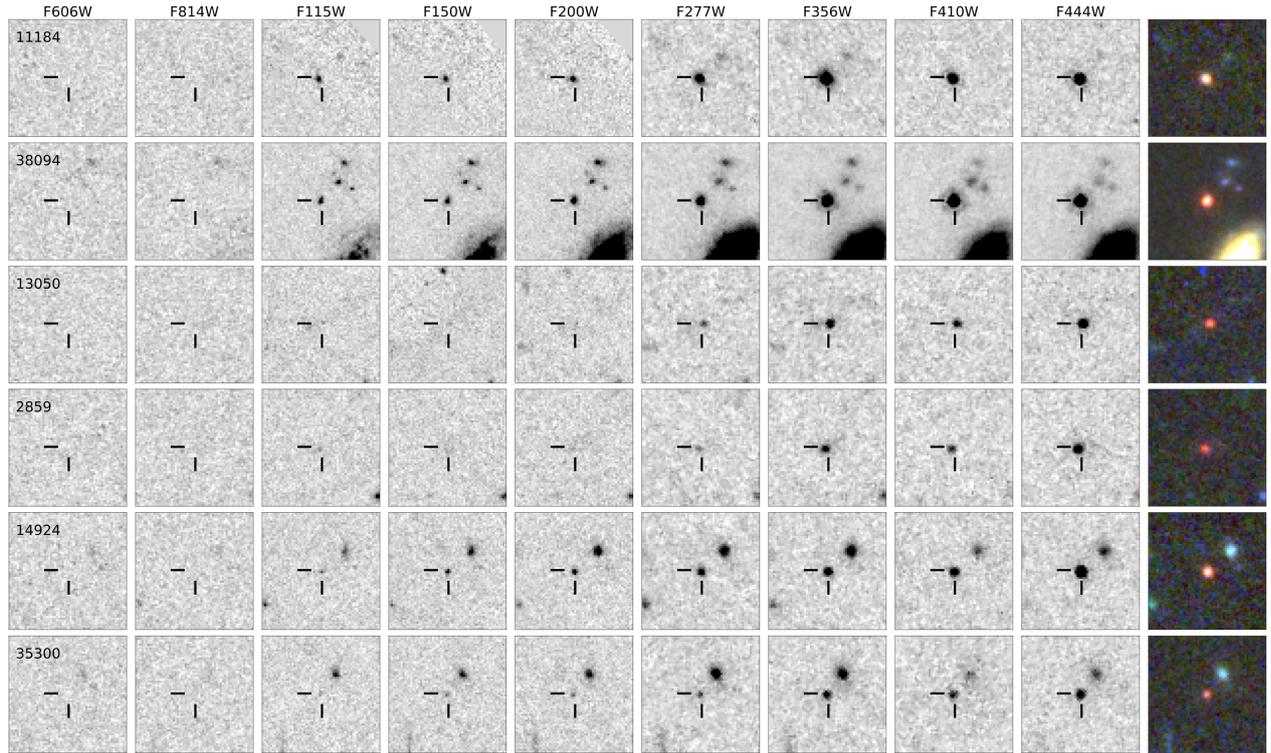

Figure 2: **Images of the six galaxies with the highest apparent masses as a function of wavelength**. The fiducial stellar masses of the galaxies are $(\log(M_*/M_\odot) > 10)$. Each cutout has a size of 2.4" × 2.4". The filters range from the 0.6 μm F606W filter of *HST*/ACS to the 4.4 μm F444W *JWST*/NIRCam filter. The galaxies are undetected in the optical filters; blue in the short-wavelength NIRCam filters; and red in the long-wavelength NIRCam filters. The color stamps show F150W in blue, F277W in green, and F444W in red.

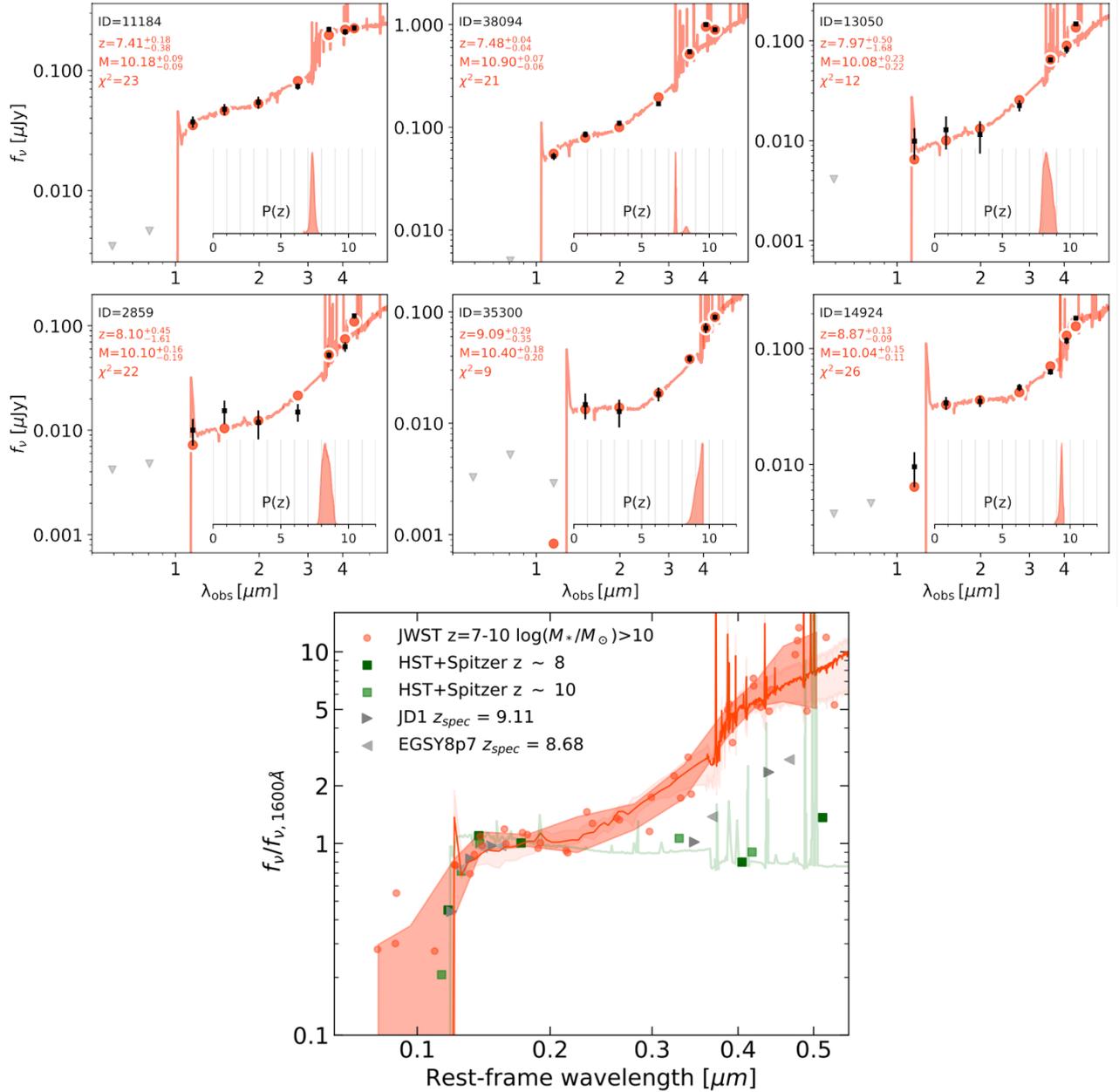

Figure 3: **Spectral energy distributions and stellar population model fits.** Top rows: photometry (black squares), best-fitting EAZY models (red lines) and redshift probability distribution P(z) (gray filled histograms) of six galaxies with apparent fiducial masses $\log(M_*/M_\odot) > 10$. The flux density units are $f_\nu$. Uncertainties and upper limits (triangles) are 1σ. Fiducial best-fit stellar masses and redshifts are noted. The seds are characterized by a double break: a Lyman break and an upturn at >3μm. Emission lines are visible in the longest wavelength bands in several cases. Bottom panel: average rest-frame SED of the 6 candidate massive galaxies (red dots) and the 16th -84th percentile of the running median (shaded area). The red line is the best-fit median EAZY model. Green squares and the green line show average rest-frame UV-selected galaxies at z=8,10 from *HST+Spitzer*[15,34]. Gray triangles show two spectroscopically confirmed galaxies at z~9[23,36,44]. The double break selected galaxies are significantly redder than previously identified objects at similar redshifts. This may be due to high M/L ratios or effects that are not included in our modeling, such as AGN or exotic lines.

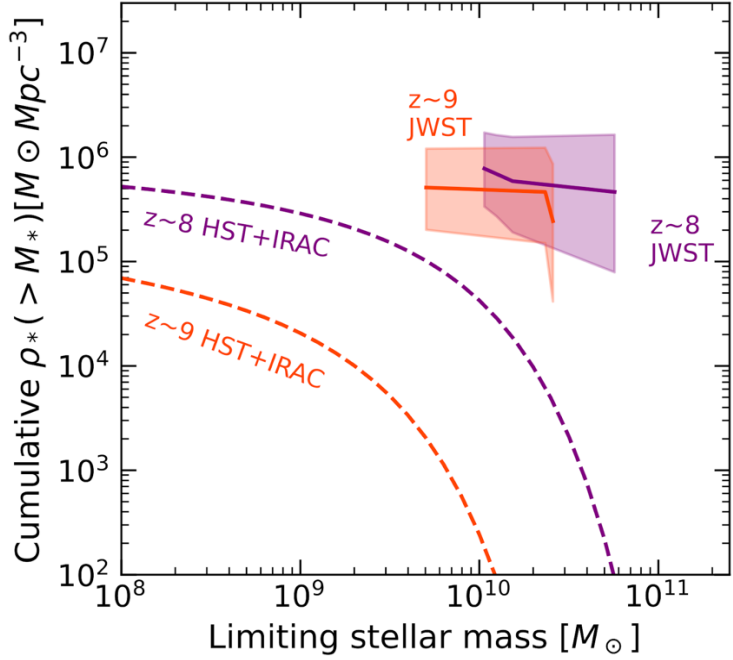

Figure 4: **Cumulative stellar mass density, if the fiducial masses of the *JWST*-selected red galaxies are confirmed.** The solid symbols show the total mass density in two redshift bins, $7 < z < 8.5$ and $8.5 < z < 10$, based on the three most massive galaxies in each bin. Uncertainties reflect Poisson statistics and cosmic variance. The dashed lines are derived from Schechter fits to UV-selected samples.[3] The *JWST*-selected galaxies would greatly exceed the mass densities of massive galaxies that were expected at these redshifts based on previous studies. This indicates that these studies were highly incomplete or that the fiducial masses are overestimated by a large factor.

**Methods**

**Observations, reduction, and photometry**

This paper is based on the first imaging taken with the Near Infrared Camera (NIRCam) on *JWST* as part of the Cosmic Evolution Early Release Science (CEERS) program (PI: Finkelstein; PID: 1345). Four pointings have been obtained, covering ~38 square arcminutes on the Extended Groth Strip *HST* legacy field and overlapping fully with the existing HST/ACS and WFC3 footprint. NIRCam observations were taken in six broadband filters, F115W, F200W, F150W, F277W, F356W, and F444W, and one medium bandwidth filter F410M. The F410M medium band sits within the F444W filter and is a sensitive tracer of emission lines, enabling improved photometric redshifts and stellar mass estimates of high-redshift galaxies[29].

Exposures produced by Stage 2 of the JWST calibration pipeline (v1.5.2) were downloaded from the MAST archive. The data reduction pipeline *Grism redshift and line analysis software for space-based spectroscopy* (Grizli[4]) was used to process, align, and co-add the exposures. The pipeline mitigates various artefacts, such as "snow-balls" and 1/f noise. the To improve pixel-to-pixel variation, custom flat-field calibration images[1] were created from on-sky commissioning data (program COM-1063) that are the median of the source-masked and background-normalized exposures in each NIRCam detector.

The pipeline then subtracts a large-scale sky background, aligns the images to stars from the Gaia DR3 catalog, and drizzles the images to a common pixel grid using astrodrizzle. The mosaics are available online as part of the v3 imaging data release[2]. Existing multi-wavelength ACS and WFC3 archival imaging from *HST* were also processed with Grizli. For the analysis in this paper all images are projected to a common 40 mas pixel grid. Remaining background structure in the NIRCam mosaics is due to scattered light. The background is generally smooth on small scales and was effectively removed with a 5" median filter after masking bright sources.

We use standard astropy[36] and photutils[37] procedures to detect sources, create segmentation maps, and perform photometry. The procedures are like those used in previous ground- and space-based imaging surveys. Briefly, we create an inverse variance weighted combined F277W + F356W + F444W image and detect sources after convolution with a Gaussian of 3 pixels FWHM (0."12) to enhance sensitivity for point sources. PSFs were matched to the F444W-band using photutils procedures. Photometry was performed at the locations of detected sources in all filters using 0."32 diameter circular apertures. The fluxes were corrected to total using the Kron autoscaling aperture measurement on the detection image. A second small correction was applied for light outside the aperture based on the encircled energy provided by the WebbPSF software. The final catalog contains 42,729 sources and includes all available HST/ACS and JWST/NIRCam filters (10 bands, spanning 0.43 to 4.4 micron). Photometry for HST/WFC3 bands was also derived, but only used for zeropoint testing as the HST/WFC3 images are several magnitudes shallower than NIRCam.

**Photometric zeropoints**

---

[1] https://s3.amazonaws.com/grizli-v2/NircamSkyflats/flats.html
[2] https://s3.amazonaws.com/grizli-v2/JwstMosaics/v3/index.html

The first *JWST* images were released with pre-flight zeropoints for the NIRCam filters. The pre-flight estimates do not match the in-flight performance, with errors up to ~20% in the long wavelength (LW) bands. This analysis uses updated in-flight calibrations that were provided by STScI on 7/29/2022 (jwst_0942.pmap) based on observations of two standard stars. The calibrations improved the accuracy of the LW photometry but introduced errors in the short wavelength (SW) bands, with variations up to 20% between detectors, as determined from comparisons to previous HST/WFC3 photometry and analyses of stars in the LMC and the globular cluster M92[38,39].

We derived new zeropoints for all SW and LW bands, for both NIRCam modules, using two independent methods. The first method ("GB") uses zeropoints that are based on standard stars observed by *JWST* in the B module and transferred to the A module using overlapping stars in the LMC. The second method ("IL") uses 5-10k galaxies at photometric redshifts 0.1 < z < 5 with SNR>15 from the CEERS parent catalog and calculates the ratio between observed and EAZY model fluxes for each detector, module, and photometric band. As the observed wavelengths sample different rest-frame parts of the SEDs of the galaxies, errors in the model fits can be separated from errors in the zeropoints. More information on the methodology and the resulting zeropoints are provided on github[3].

The methods agree very well, with differences of 3 ± 3 % in all bands except F444W, where we find a difference of 8%. We use the GB values for all bands except F444W where we take the average of the GB and IL values (multiplicative corrections 1.064 for module A and 1.084 for module B). Using the fiducial zeropoints, Extended Data Figure 1 shows offsets with respect to EAZY model fluxes, split by detector, module, and filter, showing only 0-3% residuals. A third independent method used color-magnitude diagrams of stars in M92[38,39] in F090W, F150W, F277W, F444W bands, with reported consistency with the "GB" values within the uncertainties. Our adopted zeropoints agree with the most recent NIRCam flux calibration (jwst_0989.pmap, Oct 2022) to within 4%. This paper adopts a 5% minimum systematic error (added in quadrature) for all photometric redshift and stellar population fits to account for calibration uncertainties. Finally, we compiled a sample of 450 galaxies with spectroscopic redshifts 0.2 < z < 3.8 from 3DHST[40] and MOSDEF[41] to test photometric redshift performance, finding a normalized median absolute deviation of $(z_{phot}-z_{spec})/(1+z_{spec})$ = 2.5%.

**Sample selection**

The *JWST*/NIRCam imaging in this paper reaches 5σ depths from 28.5 to 29.5 AB, representing an order of magnitude increase in sensitivity and resolution beyond wavelengths of 2.0 μm, and allowing us for the first time to select galaxies at rest-frame optical wavelengths to z ~ 10. To enable straightforward model-independent reproduction of the sample we employ a purely empirical selection of high-redshift galaxies based on NIRCam photometry, rather than one on inferred photometric redshift or stellar mass. We select on a "double break" SED: no detection in the HST ACS optical, blue in the NIRCam SW filters, and red in the NIRCam LW filters, which is expected for sources at z ≳ 7 with Lyman-break and with red UV-optical colors.

The following color selection criteria were applied:

---

[3] https://github.com/gbrammer/grizli/pull/107

$$F150W - F277W < 0.7$$
$$F277W - F444W > 1.0$$
in addition to a non-detection requirement in HST ACS imaging
$$SNR(B_{435}, V_{606}, I_{814}) < 2$$
To ensure good SNR, we limit our sample to F444W< 27 AB magnitude and F150W< 29 AB magnitude and require $SNR(F444W)>8$. We manually inspected selected sources and removed a small number of artefacts, such as hot pixels, diffraction spikes, and sources affected by residual background issues or bright neighbors.

This selection complements the traditional "drop-out" color selection techniques based on isolating the strong Lyman 1216 Å break as it moves through the filters. Drop-out selection is not feasible here: the HST ACS data are not deep enough to select dropout galaxies to the same equivalent limits as the NIRCam imaging. Screening for two breaks has shown to be an effective redshift selection: a similar technique was used to successfully select bright galaxies at 7 < z < 9 from wide-field HST and Spitzer data[29]. A red F277W-F444W color can be produced by large amounts of reddening by dust, evolved stellar populations with a Balmer Break[24], strong optical emission lines[10], or a combination of these.

This selection produced a total of 13 sources, with a median S/N ratio in the F444W band of ~ 30. The resulting sample is dark at optical wavelengths (2σ upper limit of $I_{814}$ > 30.4 AB) and faint in F115W and F150W with median ~ 28 AB magnitude, beyond the limits reached with HST/WFC3 except in small areas in the Hubble Ultra Deep Field and the Frontier Fields. The absence of any flux in the ACS optical, the red $I_{814}$ − F115W > 2.5 and blue F115W − F150W ~ 0.3 AB colors are consistent with a strong Lyman break moving beyond the ACS $I_{814}$ band at redshifts z > 6. The NIRCam F444W magnitudes are bright ~ 26 AB, and the median F150W − F444W ~ 2 AB color is redder than any sample previously reported at z > 7[3,18,21,29,42].

**Fits to the photometry**

Several methods are used to derive redshifts and stellar masses, all allowing extremely strong emission lines combined with a wide range of continuum slopes: 1) *EAZY* with additional templates that include strong emission lines, 2) *Prospector* with a strongly rising SFH prior which favors young ages, 3) *Bagpipes* to evaluate dependence on stellar population model assumptions and minimization algorithm. Finally, we also consider 4) a proposed template set for high redshift galaxies with blue continua, strong emission lines, and a non-standard IMF. Throughout, reported uncertainties are the 16th-84th percentile of the probability distributions. A Salpeter[43] IMF is assumed throughout, for consistency with previous determinations of the high redshift galaxy mass function[3,28] and constraints on the IMF in the centers of the likely descendants[44-47]. A summary of the results is presented in Extended Data Figure 3 and 4.

*1. EAZY.* The main benefits of EAZY[5] are ease of use, speed, and reproducibility. EAZY fits non-negative linear combinations of templates, with redshift and scaling of each template as free parameters. The allowed redshift range was 0 − 20 and no luminosity prior was applied. The standard EAZY template set (*tweak_fsps_QSF_12_v3*) is optimized for lower redshift galaxies. High redshift stellar populations tend to be younger, less dusty, and have stronger emission lines. We create a more appropriate template set by removing the oldest and dustiest templates ($A_V$ > 2.5) from the standard set, keeping templates 1, 2, 7, 8, 9, 10, and 11, and adding two *Flexible Stellar Population Synthesis (FSPS)* templates with strong emission lines. The first has a continuum that is approximately constant in $F_v$ with EW(Hβ+[OIII]) =

650 Å, similar to NIRSpec-confirmed galaxies[48] at z=7-8. The second has a red continuum that is constant in $F_\lambda$ with EW(H$\beta$+[OIII]) = 1100 Å, comparable to line strengths inferred for bright LBGs at z=7-9[29]. Each template has an associated M/L ratio, so the template weights in the fit can be converted to a total stellar mass. We fit all galaxies in the catalog with the default EAZY template set first and then re-fit all galaxies at z > 7 using the new template set. The template set is available online with the photometric catalog.[4] The EAZY redshift distribution of the sample of 13 galaxies is 7.3 < z < 9.4, with no low-redshift interlopers (z<6). EAZY masses range from 9.2 < log ($M_*/M_\odot$) < 10.9.

*2. Prospector.* We perform a stellar population fit with more freedom than is possible in EAZY using the Prospector[17,49] framework, specifically the Prospector-α settings[50] and the MIST stellar isochrones[51,52] from Flexible Stellar Population Synthesis (FSPS)[53,54]. This mode includes non-parametric star formation histories, with a continuity prior that disfavors large changes in the star formation rate between time bins.[55] It uses a two-component, age-dependent dust model, allows full freedom for the gas-phase and stellar metallicity, includes nebular emission where the nebulae are self-consistently powered by the stellar ionizing continuum from the model.[56] The sampling was performed using the dynesty[57] nested sampling algorithm. We also adopt two new priors which disfavor high-mass solutions: first, a mass function prior on the stellar mass, adopting the observed z=3 mass function for z>3 solutions[58], and second, a nonparametric SFH prior which favors rising SFHs in the early universe and falling SFHs in the late universe, following expectations from the cosmic star formation rate density. These are described in detail in Wang et al. (submitted).

The masses from Prospector are consistent within the uncertainties with the EAZY masses, with a mean offset of log ($M_{*Prosp}/M_{*EAZY}$)= 0.1 for objects with > $10^{10} M_\odot$. The most massive objects as indicated by EAZY are also the most massive in the Prospector fits. Prospector also provides ages and star formation rates. The star formation rates are generally not very well constrained in the fits, due to the lack of IR coverage. The ages are also uncertain and depend strongly on the adopted prior. For a constant SFH prior Prospector finds typical ages of ~0.3 Gyr, with substantial Balmer Breaks, whereas for strongly rising SFHs Prospector finds a median mass-weighted age of 34 Myr, with strong emission lines and large amounts of reddening ($A_V$ ~ 1.5). This is reminiscent of the age-dust degeneracy that is well known at lower redshift. Importantly, the stellar masses do not vary significantly between these two priors. The red SEDs (see Figure 3) require high M/L ratios for a large range of the best-fit stellar population ages, as is well known from studies of nearby galaxies[59].

*3. Bagpipes.* Fits with the Bayesian Analysis of Galaxies for Physical Inference and Parameter EStimation (Bagpipes[60]) software are also considered. Compared to Prospector, Bagpipes uses the Bruzual & Charlot stellar population models[61] and sampling algorithm Multinest[62]. While Bagpipes does not cover new parameter space compared to Prospector, it allows us to evaluate how sensitive the masses are to the adopted stellar population model or fitting technique. Furthermore, Bagpipes is relatively fast, so we can use it explore the effect of modeling assumptions to investigate the role of systematic uncertainties on the derived redshift and stellar mass. We focus on attenuation law, SFH, age sampling priors, and SNR.

A. *bagpipes_csf_salim*: baseline model of constant SFH with redshift 0 to 20, age_max from 1 Myr to 10 Gyr, metallicity between 0.01 and 2.5 Solar, ionization parameter -4 < log(U) < -2, a Salim[63] attenuation 0 < Av < 4, and adopting a linear prior in age and log prior in metallicity and ionization and uniform prior in redshift, age, and Av. The Salim law varies between a steep SMC-like extinction law at low optical depth and a flat Calzetti-like dust law at large optical depth, in accordance with empirical studies[63] and theoretical expectations[64]. The Bagpipes masses and redshifts are similar on average to those of EAZY and Prospector, with a mean offset of $\log(M_{*A}/M_{*EAZY})= 0.0$ for the massive sample.

---
[4] https://github.com/ivolabbe/red-massive-candidates

B. *Bagpipes_rising_salim:* this model is not intended to search for best fit in a wide parameter space but only in a restricted space to increase the emission line contribution to the reddest filter, F444W, and decrease the stellar masses. The model is restricted to rising star formation rates at high redshift (delayed $\tau > 0.5$ Gyr) and redshifts to $z<9.0$ to force the Hb+[OIII] complex to fall within the F444W filter. The fits show strong emission lines, low ages (median ~30 Myr) and high dust content (median $A_V \sim 1.7$). Even with these restrictions, the mean stellar mass agrees well with the baseline (mean $\log(M_{*B}/M_{*A})= -0.1$ for objects with $> 10^{10} M_\odot$).

C. *Bagpipes_csf_salim_logage*: like model (A) but with a logarithmic age prior, which is heavily weighted towards very young ages. For the 5 reddest, most massive galaxies in A the results are unchanged, whereas 6 other galaxies are now placed at significantly lower masses (inconsistent with model A, given the uncertainties), including 14924 (from $\log(M_*/M_\odot) =$ 10.1 to 8.7). The P(z) of these lower mass solutions is remarkably narrow and clustered in narrow spikes at z = 5.6, 6.9, 7.7, where the F410M filter cannot distinguish between strong lines and continuum SEDs (see Extended Data Figure 5 and 6).

D. *Bagpipes_csf_salim_logage_snr10*: to test if the fit in (C) is driven by the high SNR in long wavelength filters (which put all the weight in the fits there), we impose an error floor of 10% on the photometry which approximately balances the SNR across all NIRCam bands. Since JWST is still in early days of calibration, some limit on SNR is prudent. The SNR-limited fits result in high mass solutions for 11/13 galaxies. Notably, the uncertainties on the stellar mass do not encompass the low mass solution from (C) indicating that detailed assumptions on the treatment of SNR can introduce systematic changes.

E. *Bagpipes_csf_smc_logage*: SMC-extinction is often used in modeling high-redshift galaxies[14]. Our Bagpipes modeling use Salim-type dust which includes the SMC-like extinction at low optical depth, but it is useful to evaluate fits that are restricted to a steep extinction law in combination with a logarithmic age prior favoring young ages. The results are remarkably different from any of the modeling above: 10/13 galaxies show very low stellar masses (in the range $10^8 M_\odot$-$10^9 M_\odot$) in combination with extremely young ages (1-5 Myr). Another notable aspects are that these fits do not match the blue part of the SED well (NIRCam SW F115W, F150W, F200W) and the fits appear driven by the high SNR in the NIRCam LW filters (see Extended Data Figure 3). Most fits have significantly worse $\chi^2$ than the high-mass fits (Eazy, Prospector, Bagpipes A-D).

In conclusion, the derived masses depend on assumed attenuation law, parameterization of ages, and treatment of photometric uncertainties. Together, these aspects can produce lower redshifts and lower masses by up to factors of 100 in ways that are not reflected by the random uncertainties. Therefore, different assumptions can change the stellar masses and redshifts systematically and the uncertainties are likely underestimated.

While neither high, nor low-mass models can be excluded with the currently available data, there are two features that would suggest the ultra-young, low-mass solutions are less plausible. First, while 1-5 Myr ages are formally allowed, the galaxy would not be causally-connected – $10^{8.5} M_\odot$ of star formation would have started spontaneously on timescales less than a dynamical time (although dynamical times are uncertain until velocity dispersions and corresponding sizes are measured). In addition, the probability of catching most galaxies at that precise moment is low - given the ~200 Myr search window at z=7-9. It would suggest there are >40 older and more massive galaxies for every galaxy in our sample.

Second, the P(z) of the low-mass solutions are extremely narrow and concentrated at nearly discrete redshifts z=5.6, 6.9, 7.7 (e.g, 38094 z=6.93 +/- 0.01). Here strong Hα and Hβ+[OIII] transition between the overlapping F356W, F410M, and F444W filter edges (see Extended Data Figure 5). A single line can contribute to several bands (e.g., [OIII]5007 at z=6.9), with great flexibility due to the rapidly varying transmission at the filter edges. The result is that line and continuum dominated models are degenerate due to undersampling of the SED and resulting aliasing, but only at specific redshifts.

While finding one 5 Myr galaxy exactly in this narrow window could be luck, we find that 10/13 galaxies can only be fit with low mass, ultra-young models at these discrete redshifts z=5.6, 6.9, 7.7. Such an age and P(z) distribution for the sample, at precisely the redshifts where this fortuitous overlap between filters occurs (∼< 8% of the redshift range between z=5-9), is not implausible. To rule out that the spiked nature of the P(z) is the result of our double break selection, we perform simple simulations. We take random draws from the posteriors of line-dominated model E, redshift the models to a uniform distribution between 4 and 10, perturb with the observational errors, and apply our double break selection criterion to the simulated photometry (see Extended Data Figure 6). This suggests that even if the sample were line-dominated with ages < 5 Myr, the redshift distribution should be different (not spiked) suggesting that these fits suffer from aliasing. In contrast, P(z) of high-mass model B is broadly self-consistent with the selection function based on the model B fits. The likely reason that this effect primarily occurs with an SMC extinction law is because of the strong wavelength dependence (steep in the FUV, flatter in optical). For the sample in this paper, fits with SMC have difficulty reproducing the overall (rest-optical) red SED shape. This can be clearly seen in Extended Data Figure 3, where the SMC based fits have strongly "curved" continuum, which are generally too steep in the rest-UV and too flat in the rest-optical (F356W,F410W,F444W bands), requiring strong emission lines at specific redshifts to produce the red colors.

5. *FSPS-hot model.* For completeness we also consider recently proposed *"fsps-hot"* models[65], which consist of templates with blue continua, strong emission lines, and with a modified extremely bottom-light IMF which produces lower masses. Such an IMF is proposed to be appropriate for the extreme conditions that might be expected in high redshift galaxies. For 10 of 13 galaxies (including all massive $> 10^{10} M_\odot$ sources), the *fsps-hot* template set provides poorer fits to the photometry than the *fsps-wulturecorn* set (median $\Delta\chi^2$ = 31), due to the lack of red templates. The *fsps-hot* set places 9/13 galaxies in a narrow redshift range z=7.7 with very small uncertainties σ(z) = 0.05, reminiscent of the spiked distribution found earlier for Bagpipes model E. The blue template set can only produce red colors if strong emission lines are placed at specific redshifts. Since the fits are overall poor and no additional insight is gained, we do not consider these masses further to avoid confusion due to adopting vastly different IMFs. The extremely bottom-light IMF, with suppression of (invisible) low mass stars, is untestable with photometric data.

**Fiducial redshifts and stellar masses**

The majority of methods explored produce good fits and consistent masses and redshifts. Rather than favor one method over the others we derive "fiducial" masses and redshifts for each object by taking the median values of the EAZY (1), Prospector (2), the 5 Bagpipes fits (3-7) results of each galaxy. As discussed in the main text, the consistency between various methods may largely indicate a consistency in underlying assumptions. Different assumptions can change the stellar masses and redshifts systematically in ways that are not reflected by the random uncertainties.

Additionally, we do not consider contributions from exotic emission line species nor include AGN templates in the fits[14]. All objects in this paper should be considered candidate massive galaxies, to be confirmed with spectroscopy.

**Lensing**

A potential concern is that the fluxes (and therefore the masses) of some or all the galaxies are boosted by gravitational lensing. No galaxy is close to the expected Einstein radius of another object. The bright galaxy that is 1.2″ to the southwest of 38094 has $z_{grism} \approx 1.15$ and $M_* \approx 10.63$ (object number 28717 in 3D-HST AEGIS catalog[23]), and an Einstein radius (~ 0.4″) that is 0.3× the distance to 38094. If we assume that the mass profile of the lensing galaxy is an isothermal sphere, then the magnification is $1/(1 - \theta_E/\theta)$ where $\theta$ is the separation from the foreground source and $\theta_E$ is the Einstein radius. This would imply a relatively modest -0.15 dex correction to the stellar mass. We apply this correction when calculating densities in Figure 4.

**Volume**

Stellar mass densities for galaxies with $M_* > 10^{10}$ $M_\odot$ are calculated by grouping the galaxies in two broad redshift bins (7 < z < 8.5 and 8.5 < z < 10). At z ~ 8.5 the Lyman Break moves through the F115W filter, allowing galaxies to be separated into the two bins. The cosmic volume is estimated by integrating between the redshift limits over 38 sq arcmin, making no corrections for contamination or incompleteness. The key result is driven by the most massive galaxies. Any incompleteness would increase the derived stellar mass densities, while contamination would decrease it. Cosmic variance is about 30%, calculated using a web calculator[66,5]. The error bars on the densities are the quadratic sum of the Poisson uncertainty and cosmic variance, with the Poisson error dominant. The volume estimate is obviously simplistic, but the color selection function (see Extended Data Figure 6) suggests that most of the sample should lie between 7 < z < 10. A more refined treatment does not seem warranted given that the main (orders of magnitude) uncertainty in our study is the interpretation of the red colors of the galaxies.

**Data Availability.** The HST data are available in the Mikulski Archive for Space Telescopes (MAST; http://archive.stsci.edu), under program ID 1345. Photometry, EAZY template set, fiducial redshifts, and stellar masses of the sources presented here are available at https://github.com/ivolabbe/red-massive-candidates.

**Code Availability.** Publicly available codes and standard data reduction tools in the Python environments were used: Grizli,[4] EAZY[5], astropy[64], photutils[65], Prospector[17,37,38].

**Acknowledgements.** We are grateful to the CEERS team for providing these exquisite public JWST data so early in the mission. We thank Michael Boylan-Kolchin for helpful discussions on the theoretical context of this work. Cloud-based data processing and file storage for this work is provided by the AWS Cloud Credits for Research program. The Cosmic Dawn Center is funded by the Danish National Research Foundation. K.W. wishes to acknowledge funding from Alfred P. Sloan Foundation Grant FG-2019-12514. M.S. acknowledges project PID2019-109592GB-I00/AEI/10.13039/501100011033 from the Spanish Ministerio de Ciencia e Innovacion - Agencia Estatal de Investigacion.

**Author Contributions.** I.L. performed the photometry, devised the selection method, and led the analysis. P.v.D. drafted the main text. I.L. wrote the Methods section and produced the figures. G.B. developed the image processing pipeline and created the image mosaics. E.N. and R.B. identified the first double break galaxy, prompting the systematic search for these objects. J.L., B.W., K.S., and E.M. ran the Prospector analysis. All authors contributed to the manuscript and aided the analysis and interpretation.

**Author Information.** The authors declare that they have no competing financial interests. Correspondence and requests for materials should be addressed to I.L. (email: ilabbe@swin.edu.au).

**Extended Data**

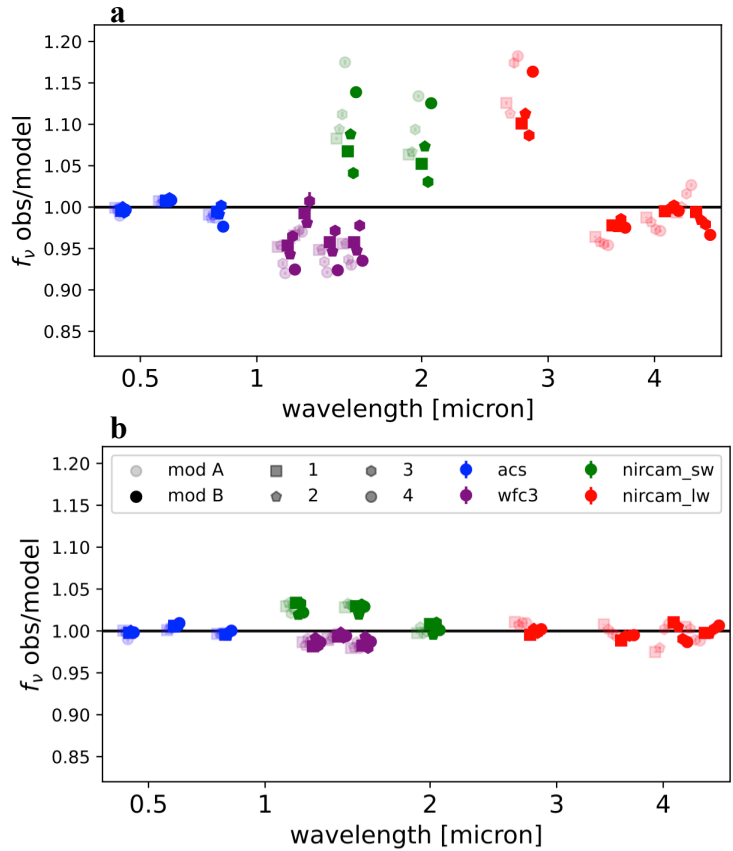

Extended Data Figure 1. **Systematic offsets in photometry as a function of wavelength.** The offsets are estimated by the ratio of the observed fluxes to the EAZY best-fit model fluxes for 5,000-10,000 sources at 0.1 < z < 5 in the CEERS field. The offsets are calculated separately for each detector (1-4), module (A/B), and filter. Symbols are slightly spread out in wavelength for clarity. **a.** the first in-flight NIRCam flux calibration update of 29 July 2022 (jwst_0942.pmap) introduced significant offsets in NIRCam short-wavelength zeropoints. **b.** After adopting our fiducial zeropoints, residual offsets are ~<3% across all bands. This paper adopts a 5% minimum systematic error for all photometric redshift and stellar population fits.

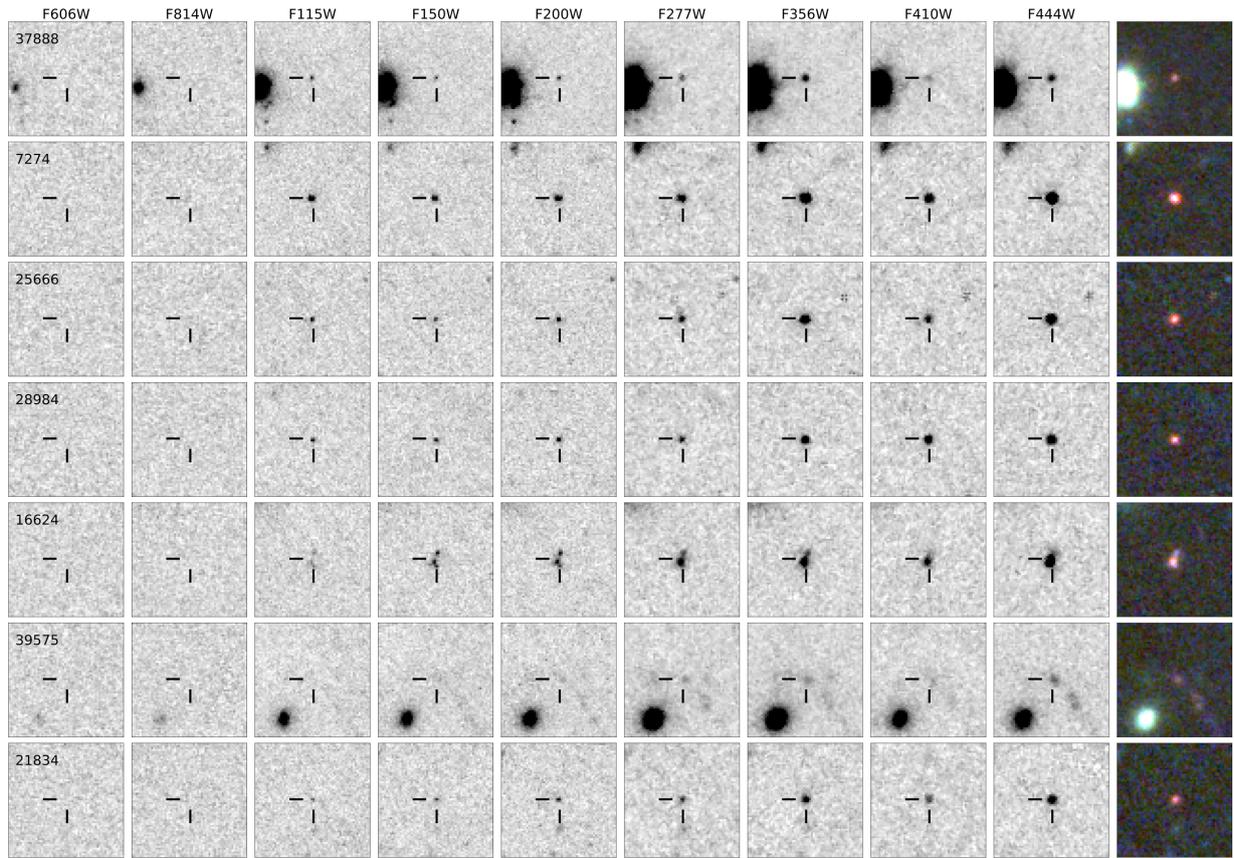

Extended Data Figure 2. **Images of the seven galaxies with apparent lowest mass.** The galaxies satisfy the color-color selection and have fiducial masses $\log(M_*/M_\odot) < 10$. The layout and panels of the figure are identical to Fig. 2 in the main text. Each cutout has a size of 2.4" × 2.4". The filters range from the 0.6 μm F606W filter of HST/ACS to the 4.4 μm F444W JWST/NIRCam filter.

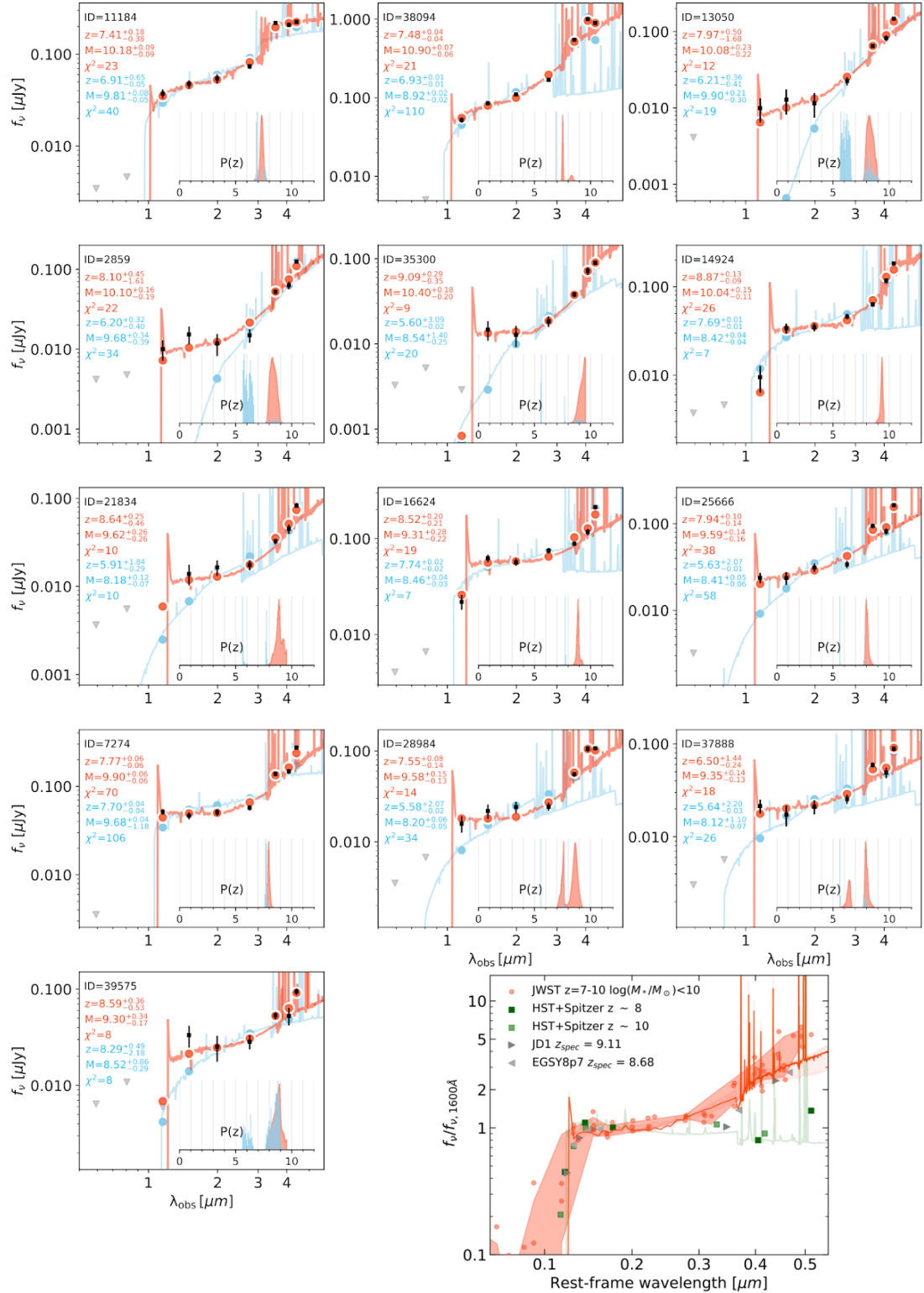

Extended Data Figure 3. **Spectral energy distributions of all 13 galaxies that satisfy the color-color selection.** The layout of the figure is identical to Fig. 3 in the main text. In addition, an alternative model fit (model E, see Methods) is shown that produces low stellar masses (blue), but generally requires extremely young ages (<5 Myr) at specific narrow redshift intervals. The panel at the lower right shows the averaged rest-frame SED of the seven galaxies with fiducial $\log(M*/M_\odot) < 10$, compared to previously-found galaxies at similar redshifts (see Fig. 3).

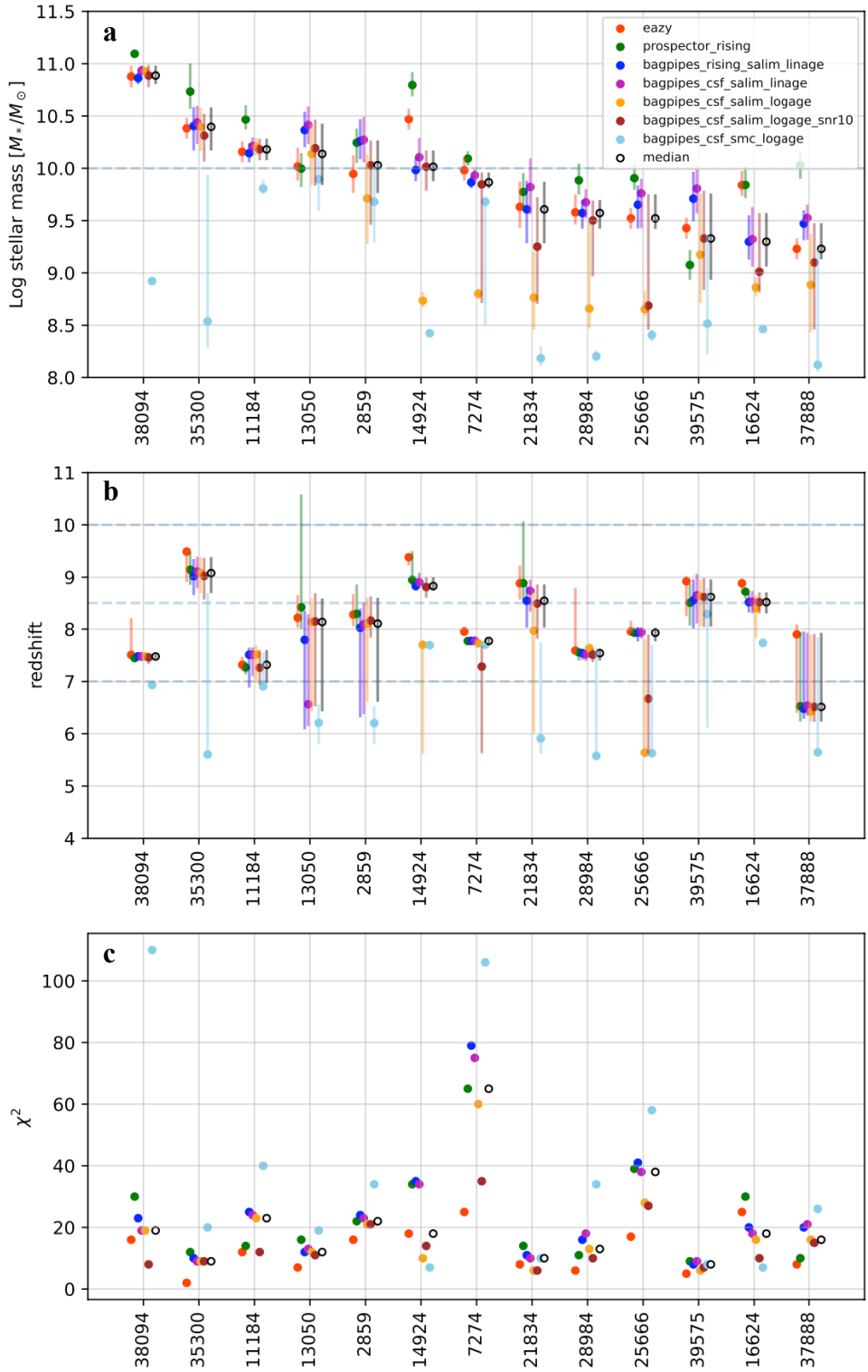

Extended Data Figure 4. **Results of the stellar population fitting.** Masses (**a**), redshifts (**b**), and the chi-squared fit quality (**c**) of the 13 galaxies that satisfy the color-color selection. For each galaxy seven different measurements are shown, as well as the median of the seven that is adopt as the fiducial value (see Methods section). These medians are listed in Table 2.

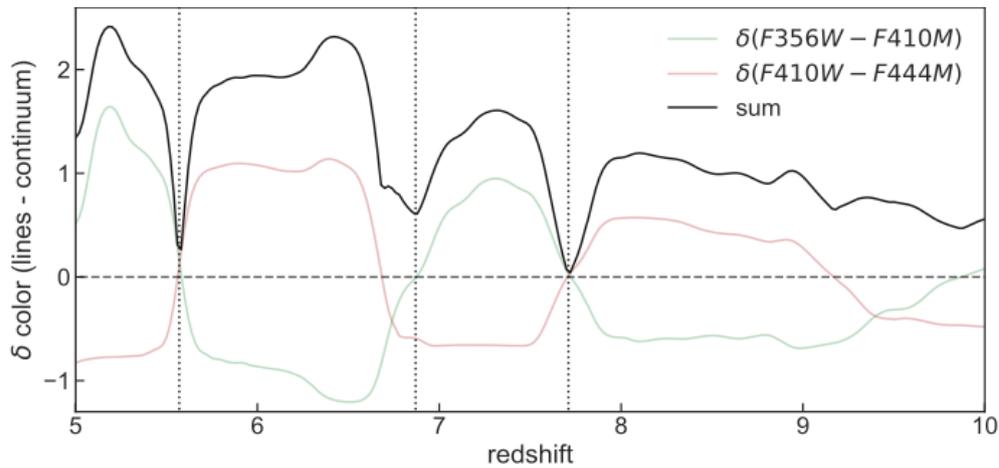

Extended Data Figure 5. **Color difference between emission line and continuum-dominated models.** The line-dominated model is a 5 Myr old constant SFH with nebular emission lines. The continuum dominated model is a 50 Myr old CSF without emission lines. Two colors differences involving the line-sensitive F410M filter are shown: F356W-F410M (green) and F410M-F444W (red) and the sum of their absolute values. When Hα and Hβ+[OIII] move through the filters with redshift, the emission line sensitive medium-band F410M filter produces a strong signature, except at z=5.6, 6.9, 7.7, where the lines transition between filters. Here continuum and line-dominated SEDs produce similar colors due to undersampling of the SED by the filters.

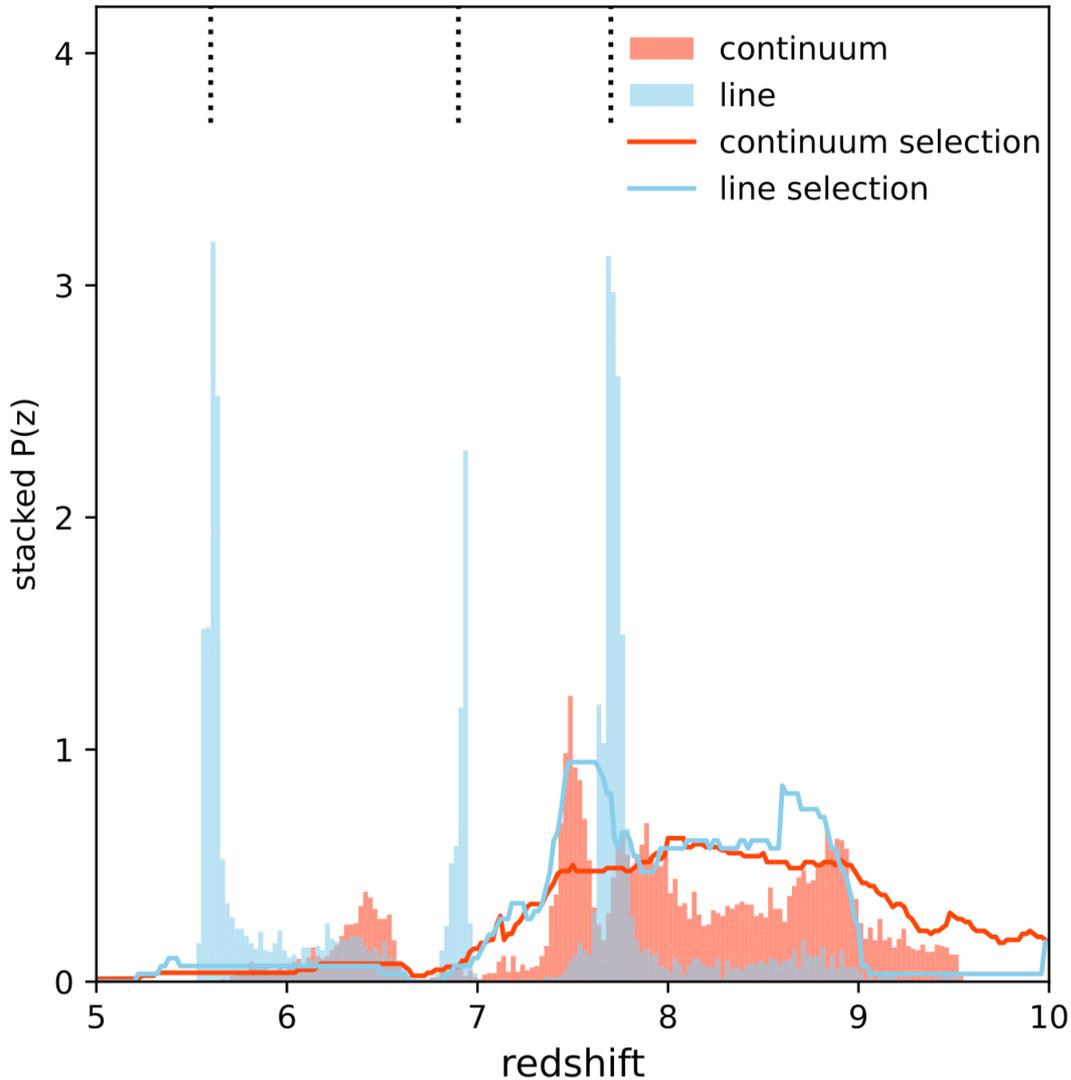

Extended Data Figure 6. **Stacked redshift probability distribution of all 13 galaxies in the sample.** The P(z) were derived using Bagpipes (as described in Methods). Redshifts of a high mass solution are shown in red (model B: Salim dust attenuation law, rising SFH, linear age prior, continuum dominated) and a low mass solution are shown in blue (model E: SMC dust, logarithmic age prior, emission line dominated). Other high mass fits (e.g., Prospector, EAZY) and low mass fits produce similar P(z). Solid curves show expected selection function under the assumption of continuum (red) or line-dominated models (blue). The high-mass continuum-dominated P(z) broadly traces the expected selection functions. The low-mass line-dominated P(z) is not expected for selection of a line-dominated model. The P(z) is concentrated at narrow redshifts around z=5.6, 6.9, 7.7 (black dotted lines) where the line-sensitive F410M cannot distinguish between continuum and strong lines due to aliasing.

Table 1. HST/ACS and JWST/NIRCam Photometry of the double break sample.

| id | f435w | f606w | f814w | f115w | f150w | f200w | f277w | f356w | f410m | f444w |
|---|---|---|---|---|---|---|---|---|---|---|
| 2859 | 3±4 | -2±4 | 4±5 | 10±3 | 15±4 | 12±4 | 15±3 | 52±2 | 63±6 | 125±3 |
| 7274 | -12±6 | 1±4 | 6±6 | 52±3 | 47±4 | 51±3 | 57±3 | 138±2 | 147±5 | 273±3 |
| 11184 | -10±6 | -4±3 | -2±5 | 37±4 | 47±4 | 54±6 | 74±3 | 219±2 | 209±6 | 225±3 |
| 13050 | 5±8 | -2±4 | 7±7 | 10±3 | 13±5 | 12±4 | 23±3 | 65±2 | 82±6 | 148±4 |
| 14924 | –±– | -3±4 | 1±5 | 10±3 | 34±4 | 35±3 | 46±2 | 63±2 | 117±5 | 183±2 |
| 16624 | –±– | 1±4 | -3±7 | 22±4 | 63±5 | 57±4 | 75±3 | 89±2 | 117±7 | 212±3 |
| 21834 | 3±4 | -1±4 | 2±6 | 4±3 | 14±4 | 17±3 | 18±2 | 33±2 | 45±5 | 83±3 |
| 25666 | -5±7 | 2±3 | 10±7 | 24±3 | 24±4 | 31±3 | 34±3 | 94±2 | 82±6 | 163±3 |
| 28984 | -3±7 | 2±4 | -1±7 | 16±3 | 22±4 | 24±3 | 24±2 | 55±2 | 105±5 | 107±3 |
| 35300 | –±– | -4±3 | 4±5 | 1±3 | 15±4 | 13±4 | 18±2 | 38±2 | 72±7 | 90±3 |
| 37888 | 1±5 | -4±3 | 2±6 | 21±4 | 17±4 | 21±4 | 26±3 | 59±2 | 49±6 | 89±3 |
| 38094 | 2±4 | 2±4 | -6±5 | 52±3 | 86±4 | 110±3 | 169±3 | 546±3 | 1003±8 | 893±4 |
| 39575 | 3±8 | 4±6 | -6±11 | 0±6 | 33±8 | 25±8 | 28±4 | 53±4 | 53±11 | 94±6 |

Note. — Units are nJy. A fixed 5% uncertainty is added in quadrature to the photometric uncertainties account for calibration errors before fitting with `EAZY, Prospector`, and `Bagpipes`.

Table 2. Redshift and stellar masses of the double break sample.

| id | ra | dec | redshift | stellar mass $\log(M_*/M_\odot)$ |
|---|---|---|---|---|
| 2859 | 214.840534 | 52.817942 | 8.11(+0.49, −1.49)(+0.75, −2.30) | 10.03(+0.24, −0.27)(+0.46, −0.75) |
| 7274 | 214.806671 | 52.837802 | 7.77(+0.05, −0.06)(+0.27, −2.15) | 9.87(+0.09, −0.06)(+0.30, −1.36) |
| 11184 | 214.892475 | 52.856892 | 7.32(+0.28, −0.35)(+0.38, −0.46) | 10.18(+0.10, −0.10)(+0.42, −0.43) |
| 13050 | 214.809155 | 52.868481 | 8.14(+0.45, −1.71)(+2.45, −2.33) | 10.14(+0.29, −0.30)(+0.45, −0.54) |
| 14924 | 214.876150 | 52.880833 | 8.83(+0.17, −0.09)(+0.67, −3.22) | 10.02(+0.16, −0.14)(+0.90, −1.63) |
| 16624 | 214.844772 | 52.892108 | 8.52(+0.19, −0.22)(+0.46, −0.80) | 9.30(+0.27, −0.24)(+0.72, −0.87) |
| 21834 | 214.902227 | 52.939370 | 8.54(+0.32, −0.51)(+1.52, −2.92) | 9.61(+0.26, −0.32)(+0.49, −1.50) |
| 25666 | 214.956837 | 52.973153 | 7.93(+0.09, −0.16)(+0.23, −2.32) | 9.52(+0.23, −0.10)(+0.52, −1.17) |
| 28984 | 215.002843 | 53.007594 | 7.54(+0.08, −0.14)(+1.25, −1.98) | 9.57(+0.13, −0.15)(+0.47, −1.42) |
| 35300 | 214.830662 | 52.887777 | 9.08(+0.31, −0.38)(+0.40, −3.50) | 10.40(+0.19, −0.23)(+0.60, −2.11) |
| 37888 | 214.912510 | 52.949435 | 6.51(+1.42, −0.28)(+1.58, −0.90) | 9.23(+0.25, −0.10)(+0.92, −1.17) |
| 38094 | 214.983019 | 52.955999 | 7.48(+0.04, −0.04)(+0.74, −0.56) | 10.89(+0.09, −0.08)(+0.22, −1.99) |
| 39575 | 215.005400 | 52.996706 | 8.62(+0.34, −0.57)(+0.45, −2.51) | 9.33(+0.43, −0.39)(+0.69, −1.11) |

Note. — The adopted redshift and stellar mass are the medians of redshifts and masses computed with 7 different methods (`EAZY, Prospector`, and `Bagpipes`(5 variations, including dust, SFH, age prior, and SNR limit), see Methods. A Salpeter IMF is assumed. Two uncertainties are listed ($\pm(ran) \pm (sys)$) with random uncertainties (ran) corresponding to the 16th and 84th percentile median posterior distributions, and systematic uncertainties (sys) corresponding to the extremes of all model fits.